\documentclass[reprint,amsmath,amssymb,apr,aip,onecolumn,nofootinbib]{revtex4-2}
\usepackage{graphicx}
\usepackage{graphics}
\usepackage[T1]{fontenc}
\usepackage{mathptmx}
\usepackage{times}
\usepackage{amsmath}
\usepackage{amssymb}
\usepackage{dcolumn}
\usepackage{color}
\usepackage{bm}
\usepackage{units}
\usepackage{booktabs}
\usepackage{footnote}
\usepackage{multirow}
\usepackage{tabularray}
\UseTblrLibrary{booktabs}
\usepackage{threeparttable}
\usepackage[strict]{changepage}
\usepackage{algorithm}
\usepackage{algpseudocode}
\usepackage{framed}
\usepackage{makecell}

\usepackage{tcolorbox}
\tcbset{
  colback=white,
  colframe=black!15,
  boxrule=0.5pt,
  sharp corners,
  left=8pt,right=8pt,top=8pt,bottom=8pt
}

\usepackage{enumitem}
\setlist[description]{%
  font=\bfseries,
  labelwidth=4.0cm,   % increased for wider labels
  labelsep=1.2em,     % more space between label and content
  leftmargin=!,
  style=nextline       % label on same line, no auto colon duplication
}

\definecolor{formalshade}{rgb}{0.95,0.95,1}

\usepackage{acronym}
\acrodef{ADC}[ADC]{Analog-to-Digital Converter}
\acrodef{AI}[AI]{Artificial Intelligence}
\acrodef{aimc}[AIMC]{Analog In-Memory Computing}
\acrodef{ART}[ART]{Adversarial Robustness Toolbox}
\acrodef{ASR}[ASR]{Adversarial Success Rate}
\acrodef{CMOS}[CMOS]{Complementary Metal-Oxide Semiconductor}
\acrodef{CNN}[CNN]{Convolutional Neural Network}
\acrodef{DAC}[DAC]{Digital-to-Analog Converter}
\acrodef{dma}[DMA]{Direct Memory Access}
\acrodef{DL}[DL]{Deep Learning}
\acrodef{DNN}[DNN]{Deep Neural Network}
\acrodef{DSE}[DSE]{Design Space Exploration}
\acrodef{FGSM}[FGSM]{Fast Gradient Sign Method}
\acrodef{fp}[FP]{Floating-Point}
\acrodef{fpu}[FPU]{Floating-Point Unit}
\acrodef{GBDA}[GBDA]{Gradient-based Distributional Attack}
\acrodef{DPU}[DPU]{Digital Processing Unit}
\acrodef{GDP}[GDP]{Gradient Descent Programming}
\acrodef{GPU}[GPU]{Graphics Processor Unit}
\acrodef{hwa}[AHWA]{Analog Hardware-Aware}
\acrodef{IMC}[IMC]{In-Memory Computing}
\acrodef{IoT}[IoT]{Internet-of-Things}
\acrodef{LDPU}[LDPU]{Local Digital Processing Unit}
\acrodef{lora}[LoRA]{Low-Rank Adaptation}
\acrodef{MAC}[MAC]{Multiply and Accumulate}
\acrodef{ML}[ML]{Machine Learning}
\acrodef{MVM}[MVM]{Matrix-Vector Multiplication}
\acrodef{NLP}[NLP]{Natural Language Processing}
\acrodef{NN}[NN]{Neural Network}
\acrodef{NVM}[NVM]{Non-Volatile Memory}
\acrodef{pcm}[PCM]{Phase Change Memory}
\acrodef{PGD}[PGD]{Projected Gradient Descent}
\acrodef{PIM}[PIM]{Processing-In-Memory}
\acrodef{pmca}[PMCA]{Programmable Multi-Core Accelerator}
\acrodef{PTQ}[PTQ]{Post Training Quantization}
\acrodef{RGB}[RGB]{Red Green Blue}
\acrodef{RRAM}[RRAM]{Resistive Random-Access Memory}
\acrodef{llm}[LLM]{Large-Language Model}
\acrodef{rtl}[RTL]{Register-Transfer Level}
\acrodef{RTN}[RTN]{Random Telegraph Noise}
\acrodef{tcdm}[TCDM]{Tightly Coupled Data Memory}
\acrodef{FFN}[FFN]{Feed-Forward Network}
\acrodef{GRPO}[GRPO]{Group Relative Policy Optimization}
\acrodef{LSTM}[LSTM]{Long Short-Term Memory}
\renewcommand{\thefootnote}{\alph{footnote}}
\newcommand{\ignore}[1]{}

\usepackage{hyperref}
\begin{document}
\title{Efficient transformer adaptation for analog in-memory computing via low-rank adapters}

\author{Chen Li\textsuperscript{$\ddagger$}}
\affiliation{\textit{Department of Engineering, King's College London, United Kingdom}}
\author{Elena Ferro\textsuperscript{$\ddagger$}}
\affiliation{IBM Research Europe, 8803 R\"{u}schlikon, Switzerland}
\author{Corey Lammie}
\affiliation{IBM Research Europe, 8803 R\"{u}schlikon, Switzerland}
\author{Manuel Le Gallo}
\affiliation{IBM Research Europe, 8803 R\"{u}schlikon, Switzerland}
\author{Irem Boybat}
\affiliation{IBM Research Europe, 8803 R\"{u}schlikon, Switzerland}
\author{Bipin Rajendran\textsuperscript{$\ast$}}
\affiliation{\textit{Department of Engineering, King's College London, United Kingdom}}

\newcommand{\firstpagefootnote}{
    \renewcommand{\thefootnote}{\textsuperscript{$\ddagger$}}\footnotetext{These authors contributed equally.}
    \renewcommand{\thefootnote}{\textsuperscript{$\ast$}}
    \footnotetext{Corresponding author. Email: bipin.rajendran@kcl.ac.uk}
}

\firstpagefootnote

\begin{abstract}
Analog In-Memory Computing (AIMC) offers a promising solution to the von Neumann bottleneck. However, deploying transformer models on AIMC remains challenging due to their inherent need for flexibility and adaptability across diverse tasks. For the benefits of AIMC to be fully realized, weights of static vector-matrix multiplications must be mapped and programmed to analog devices in a weight-stationary manner. This poses two challenges for adapting a base network to hardware and downstream tasks: (i) conventional analog hardware-aware (AHWA) training requires retraining the entire model, and (ii) reprogramming analog devices is both time- and energy-intensive. To address these issues, we propose Analog Hardware-Aware Low-Rank Adaptation (AHWA-LoRA) training, a novel approach for efficiently adapting transformers to AIMC hardware. AHWA-LoRA training keeps the analog weights fixed as meta-weights and introduces lightweight external LoRA modules for both hardware and task adaptation. We validate AHWA-LoRA training on SQuAD v1.1 and the GLUE benchmark, demonstrate its scalability to larger models, and show its effectiveness in instruction tuning and reinforcement learning. We further evaluate a practical deployment scenario that balances AIMC tile latency with digital LoRA processing using optimized pipeline strategies, with RISC-V-based programmable multi-core accelerators.
This hybrid architecture achieves efficient transformer inference with only a \unit[4]{\%} per-layer overhead compared to a fully AIMC implementation.  Code is available$\footnote{\url{https://github.com/chenlicodebank/lora_on_analog_hardware}}$.
\end{abstract}
\maketitle

\acresetall
% \section*{Introduction}
Deep learning has revolutionized various fields, from computer vision to \ac{NLP}, achieving unprecedented performance in the solution of complex tasks \cite{lecun}.
Much of this progress stems from scaling up \ac{NN} architectures and datasets, enabling models to capture more information and approximate complex functions that can be generalized to unseen samples. Among these advancements, the transformer architecture stands out as the backbone of \acp{llm}, allowing more effective network scaling and data compression \cite{Vaswani17}. However, as \acp{NN} grow in size and complexity, they demand more computational resources, leading to higher power consumption and carbon emissions \cite{patterson2021carbon}. This raises sustainability concerns and drives research into more energy-efficient architectures tailored to \ac{NN} computation.

\ac{aimc} has emerged as a promising computing paradigm to tackle these challenges, offering improved performance and energy-efficiency through computation directly within the memory array~\cite{Sebastian17_2, Boybat_IEDM2024}.
However, distinct properties of \ac{aimc}, namely device noise and circuit non-idealities, introduce additional complexity to \ac{NN} training and deployment~\cite{Lammie2025a}.
Analog devices are inherently non-deterministic and subject to temporal variations, impacting \ac{NN} accuracy when deployed on \ac{aimc}-based accelerators~\cite{8964852,Sebastian2020-la,9371990,9720519}.
\ac{hwa} training techniques have been demonstrated to enhance model robustness under these constraints for various \ac{NN} architectures, effectively mitigating accuracy losses by injecting Gaussian noise during forward-propagation and simulating circuit-non-idealities~\cite{10.1063/5.0168089,rasch2023hardware,Lammie2025}.
However, transformer-based architectures introduce several unique challenges that make the direct application of \ac{hwa} training techniques difficult. We summarize these limitations below, which motivate the development of a novel approach to efficiently adapt transformers for \ac{aimc} hardware.

First, training large-scale networks with a high parameter count quickly becomes computationally challenging. Transformers, in particular, require significantly more parameters than traditional architectures like \acp{CNN} and \ac{LSTM} that have been the primary focus of \ac{aimc} research. While the larger size of transformers contributes to their strong performance across many tasks, it also makes \ac{hwa} training challenging. Training such large models with simulated hardware constraints using existing \ac{hwa} approaches often exceeds \ac{GPU} memory limits and results in prohibitive computational costs.

Second, while pre-trained transformer models generalize remarkably well across a wide range of natural language processing tasks due to extensive training on large and diverse corpora \cite{devlin2018bert}, conventional \ac{hwa} training methodologies typically optimize performance for only one task at a time~\cite{rasch2023hardware}. Such narrowly-focused, task-specific tuning under-utilizes the generalization potential of pre-trained models, resulting in hardware-optimized models that perform well on one task but struggle to generalize to others. While aggregating multiple tasks into a single unified corpus and conducting \ac{hwa} training on it might seem like a remedy, it often leads to degraded performance, as conflicting task objectives can interfere with the model's learned representations.

Another critical limitation is that existing \ac{hwa} methods are not designed for continual adaptation -- a necessity for many real-world applications. In dynamic environments, transformers must adapt to new data and shifting user needs. However, current \ac{hwa} approaches achieve adaptation by reloading and retraining the full model weights on \ac{aimc} hardware -- a process that is both time-consuming and energy-intensive, making frequent updates impractical~\cite{benmeziane2024multitask}.

Finally, real-world \ac{aimc}-based accelerators suffer from various device noise and circuit non-idealities. Industry-fabricated \ac{pcm} devices, for example, display state-dependent programming and read noise, as well as drift in conductance states over time~\cite{8964852,Sebastian2020-la,9371990,9720519}. While \ac{hwa} training has proven adept at addressing these hardware imperfections, a more advanced training methodology should preserve this strength while addressing the aforementioned limitations.

\begin{figure*}[!t]
\centering
\includegraphics[width=1\columnwidth]{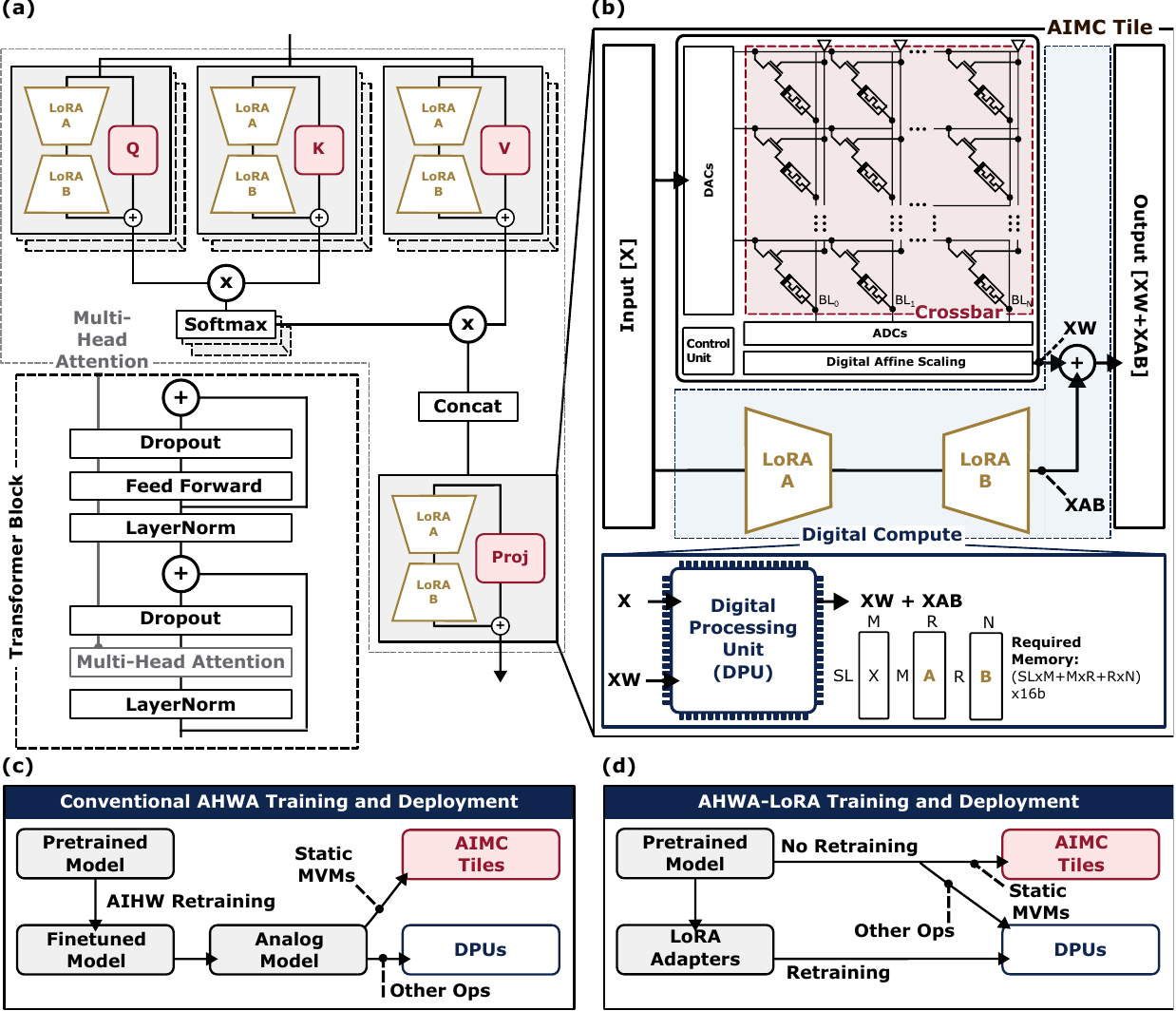} 
\caption{\textbf{Implementation of \ac{hwa}-\ac{lora} training to the multi-head attention block in the standard transformer architecture}.
\textbf{a} Within each transformer block, the weights of linear (dense) layers with fixed weights (\( W \in \mathbb{R}^{m \times n} \)) are mapped to \ac{aimc} tiles.
The \ac{lora} weight matrices \( A \in \mathbb{R}^{m \times r} \) and \( B \in \mathbb{R}^{r \times n} \) can be trained to adapt the effective weights for different downstream tasks without altering the meta-weights \( W \in \mathbb{R}^{m \times n} \).
\textbf{b} The high-level proposed architecture, where each \ac{aimc} tile is paired with a \ac{pmca}.
The latency of \ac{aimc} tiles and \acp{pmca} is balanced. Computation of \( XAB \) is performed within the \ac{pmca}, alongside the required addition operation.
\textbf{c} Conventional \ac{hwa} training and deployment methods require retraining of the meta-weights of the pre-trained model.
\textbf{d} \ac{hwa}-\ac{lora} training preserves the meta-weights from the pre-trained model and deploys them directly to the \ac{aimc} hardware. Only the \ac{lora} weight matrices are trained and executed by the \acp{DPU}.
}
\label{fig1}
\end{figure*}

Hence, we introduce \emph{\ac{hwa}-LoRA training} -- a method that leverages the principles of \ac{lora}~\cite{hu2021loralowrankadaptationlarge} to address the challenges faced by \ac{aimc} systems~\cite{hu2022lora}.
As shown in Figure \ref{fig1}, \emph{\ac{hwa}-LoRA training} preserves the original, pre-trained weights of the transformer model, i.e., the meta-weights. These meta-weights are obtained through extensive pretraining and possess strong generalization capabilities. In conventional \ac{hwa} training, these meta-weights are overwritten during task-specific fine-tuning. We argue that these meta-weights should be preserved and instead introduce lightweight \ac{lora} adapters to alter the effective weight values for different tasks.
For task-specific retraining, this approach drastically reduces the memory overhead of \ac{hwa} training.
During inference, the inclusion of \ac{lora} modules equips \ac{aimc} with on-chip adaptation ability, allowing models to dynamically respond to hardware constraints and evolving task demands. This flexibility is difficult to achieve using conventional methods.

The complete \emph{\ac{hwa}-LoRA training} pipeline consists of three main steps.
First, pre-trained transformer weights are directly mapped onto \ac{aimc} hardware (meta weight deployment). Second, hardware constraints are applied to the meta-weights, but only the LoRA weights are updated (\emph{\ac{hwa}-LoRA training}). Third, the trained LoRA weights are deployed onto \acfp{DPU}, allowing them to operate in parallel with analog computation on the meta-weights (LoRA weight deployment). A detailed description of these steps is provided in the Methods.
We investigate the accuracy of our method using a realistic heterogeneous \ac{aimc}-based hardware configuration and statistical \ac{pcm} device model calibrated on hardware measurements from a chip containing one million \ac{pcm} devices.

Simulation results confirm that our method is effective on a 25.3M-parameter transformer model, a practical size suitable for deployment on currently available \ac{aimc} chips~\cite{Le_Gallo_2023, Wen2024}.
We demonstrate the scalability of our method to larger encoder-only transformer architectures of up to approximately 300 million parameters, including BERT-Base and BERT-Large, which may be supported by future \ac{aimc} chips. Results indicate our method constrains the size of the \ac{lora} components at just ~1\% of the total model parameters. Furthermore, we extend our method to even larger decoder-only transformers, successfully applying it to the LLaMA 3.1 8B model for tasks in instruction tuning and mathematical reasoning, through supervised learning and reinforcement learning, respectively. Finally, we explore an optimized implementation that integrates \ac{aimc} computation with digital LoRA adaptation using RISC-V-based \acfp{pmca} as \acp{DPU}. By strategically partitioning the workload, i.e., assigning the meta-weights to \ac{aimc} tiles and mapping LoRA layers onto digital \ac{pmca} units, our design offers a practical and efficient framework for serving transformer models onto these \ac{aimc} chips~\cite{8268339}.

\section*{Methods}\label{sec:methods}
\subsection*{AHWA-LoRA Training}
The complete \emph{\ac{hwa}-\ac{lora} training} pipeline consisted of three main steps.
During the first step, the meta weights of the model were directly deployed to \ac{aimc} hardware without any training. The hardware-specific parameters, such as weight clipping thresholds and the bit resolutions of \ac{aimc} peripherals (e.g., \acp{ADC} and \acp{DAC}), were determined and used to simulate hardware constraints in the subsequent training process. In the second step, hardware constraints were simulated by incorporating them into the forward pass of the meta weights. Gradients propagated back through these simulated constraints, but learning happened in \ac{lora} and only the \ac{lora} weights were updated. This structure allowed the meta-weights to "sense" the hardware limitations, while \ac{lora} learned to compensate for them. Importantly, the model was trained using data from the downstream task, ensuring that it was optimized for the target application. This enabled joint optimization for both hardware compatibility and task performance via the introduced \ac{lora} component. The loss function remained unchanged from standard \ac{hwa} training; no modification was needed to incentivise this behavior. Finally, the trained \ac{lora} weights were deployed onto \acp{DPU}. These \ac{lora} weights, though high in precision, were relatively small in number, making it feasible to parallelize their computation with the analog computation performed on the fixed meta-weights residing in \ac{aimc} tiles.

In \ac{hwa}-\ac{lora} training, the meta-weights, which are the pretrained base-model weights, are programmed onto \ac{aimc} hardware once and then kept fixed during adaptation. This choice improves long-term reusability: the same analog-programmed base model can be reused across multiple downstream tasks by swapping lightweight LoRA adapters, avoiding costly and frequent array reprogramming. Moreover, by not updating analog-mapped weights, \ac{hwa}-LoRA training avoids error accumulation commonly associated with repeated programming or on-chip learning. Temporal non-idealities after deployment (e.g., conductance drift) are mitigated via global drift compensation; when necessary, LoRA adapters can be refreshed off-chip and redeployed without reprogramming the \ac{aimc} arrays.

\subsection*{Model Mapping to Hardware and Assumed Hardware Configuration}
Our approach mapped all linear layers of MobileBERT (comprising 20.4M parameters, $\sim$81\% of the total parameters) onto \ac{aimc} tiles. This included the embedding transformation layer, the final output layer, and the linear layers within both \ac{FFN} and the QKV projection of multi-head attention. The \ac{aimc} tiles had $512 \times 512$ unit cells, and 8-bit \acp{DAC} and \acp{ADC}. A digital affine scaling operation was applied after the \acp{ADC}. Due to the dynamic nature of matrix-matrix operations, the computation of attention scores was handled by \acp{pmca}, as non-volatile memory-based \ac{aimc} computation was not well-suited for such operations. Additionally, \acp{pmca} managed the LoRA adapters and integrated their outputs with those from the \ac{aimc} tiles.
To accurately model the behavior and constraints of \ac{aimc} hardware, we used AIHWKIT, an open-source simulator for \ac{aimc} devices \cite{aihwkit}. AIHWKIT offered detailed models of \ac{pcm} devices based on extensive experimental data and accurately modeled the peripheral circuitry. A differential channel-wise weight mapping scheme was applied to all analog weights with maximum conductance $G_{max}=25\mu S$, and each channel was clipped to 3-sigma based on the fitted weight distribution.

\subsection*{Training and Inference Details}
During forward propagations, we injected hardware constraints on the model, including Gaussian noise \cite{Le_Gallo_2023} with an amplitude of 6.7\% on analog weights and 4.0\% on \acp{ADC}, among other hardware constraints. The 6.7\% weight-noise level is not an immutable physical constant of \ac{pcm} devices; rather, it is an \emph{effective} noise amplitude used in our simulator to approximate the dominant stochastic behavior of the AIHWKIT \ac{pcm} analog tile model \cite{aihwkit}, which is calibrated to measured device statistics and includes multiple non-idealities and parameters. For training efficiency, we summarize these effects using a zero-mean Gaussian perturbation model as a first-order approximation. Importantly, this amplitude is a tunable hyperparameter in our framework; we provide an ablation over noise amplitudes in the Supplementary Information (Table~II), and select 6.7\% as a strong accuracy--cost trade-off under our training budget.

Noise is ``injected'' on-the-fly during each forward pass by forming a temporary noisy instance of the fixed meta-weights. Concretely, we maintain clean master meta-weights \(W\) throughout training, and for each minibatch we sample an i.i.d.\ perturbation and compute \(\tilde{W}=W+\Delta W\), where \(\Delta W\) follows a zero-mean Gaussian distribution with relative amplitude set by the chosen noise level (e.g., 6.7\%). The noisy weights \(\tilde{W}\) are used only for the current forward/backward computation and are discarded afterwards, ensuring that noise is uncorrelated across minibatches and that the expected noisy weights remain unbiased around \(W\). In \ac{hwa}-\ac{lora} training, gradients flow through the noisy forward path, but only the \ac{lora} parameters are updated, while the meta-weights remain fixed.
The Adam optimizer was used with an initial learning rate of $2 \times 10^{-4}$, following a linear decay schedule over 15 epochs. The maximum Sequence Length (SL) was set to 320.
For inference, we evaluated accuracy and robustness under realistic hardware constraints, including programming noise, conductance drift, and read noise for a period ranging from 0 seconds to 10 years. All results were averaged over 10 trials. We used global drift compensation \cite{Rajendran75} to mitigate temporal variations. We evaluated the impact of different rank values \( r \) on performance and computational overhead, determining that a rank of 8 provided an optimal balance (see Figure \ref{fig3}a).

\subsection*{Instruction Tuning and Reinforcement Learning}
For instruction tuning, we utilized a carefully curated version of the Alpaca dataset \cite{alpaca}, which enhanced data quality and consistency. Our experiments were conducted using the LLaMA-3.1 8B model, incorporating a \ac{lora} rank of 16. This corresponded to approximately 0.52\% of the model's total parameters. The model mapping strategy was consistent with the method used in MobileBERT. All linear layers were mapped to \ac{aimc} tiles and attention was managed by \acp{pmca}. During forward propagation, hardware constraints were emulated by injecting Gaussian noise with an amplitude of 6.7\% by adding it to the meta weights, alongside additional hardware-induced limitations. Explicit modeling of \ac{ADC} and \ac{DAC} components was omitted, assuming these components operated at high resolution. Unlike the approach taken in MobileBERT experiments, we did not apply weight clipping, following recent insights from weight quantization research in large language models \cite{dettmers2023qlora, frantar2022gptq}. Training was executed with the AdamW optimizer, using an initial learning rate of $2 \times 10^{-4}$, a linear decay schedule over 6,470 steps, and a brief warm-up period of 5 steps. The batch size and weight decay were set to 8 and 0.01, respectively. The maximum sequence length was limited to 2,048 tokens.

In the reinforcement learning setup, we leveraged the GSM8K dataset to foster the model's reasoning abilities, particularly encouraging the effective use of long Chain-of-Thought (CoT). We employed the \ac{GRPO} algorithm \cite{guo2025deepseek} for training, with the instruction-tuned Llama-3.1 8B model again featuring a \ac{lora} rank of 16. Consistent with our earlier instruction-tuning approach, we applied the same model mapping strategy and refrained from applying weight clipping. However, the amplitude of Gaussian noise injected during RL training was reduced to 3.0\%. {We emphasize that this reduction does not imply that the underlying \ac{pcm} noise is physically reduced; rather, it is a practical training choice for stable and compute-efficient reinforcement learning. In GRPO, the update relies on \emph{relative} quality differences among multiple sampled completions for the same prompt to form effective advantage estimates. With 6.7\% noise, the initial policy frequently produces uniformly low-quality (near-random) generations, yielding near-zero or non-informative reward differences within a sampling group and substantially weakening the learning signal. Using 3.0\% preserves a meaningful gradient signal early in training while still operating in a noisy analog regime. A straightforward way to enable higher-noise RL is a curriculum strategy (e.g., supervised adaptation at the target noise level before any reinforcement learning), which we leave for future work. Modeling of \ac{ADC} and \ac{DAC} components was reserved for future investigation. The maximum sequence length during RL training was 1,024 tokens. Our RL approach incorporated four complementary reward functions to incentivize correct reasoning patterns and precise mathematical answers, allowing a maximum achievable reward of 9.5. For each GSM8K question, we generated 16 sample outputs, which were grouped and utilized for advantage calculations by \ac{GRPO} \cite{guo2025deepseek}. RL training consisted of 500 optimization steps, including a 50-step warm-up phase, with a learning rate set at $5 \times 10^{-6}$ and a comparatively higher weight decay of 0.1.

To efficiently handle memory overhead during training, we adopted the aihwkit-lightning framework \cite{aihwkitlightning}. Its low memory footprint, combined with \ac{lora}, allowed us to perform all training and evaluations on a single \ac{GPU} with 80 GB of memory. During inference, we evaluated model robustness against Gaussian noise applied to weights, following methodologies established in prior studies \cite{Rajendran75, buchel2025analog}. The Gaussian noise levels during evaluation were set to match those used during training: 6.7\% for instruction tuning and 3.0\% for reinforcement learning. Additional evaluation results for a wider range of noise levels, as well as results using the \ac{pcm} model with zero-second drift, were provided in the Supplementary Information.

\subsection*{{PMCA Performance Estimation Simulations}}
We adopted a small version of the Snitch cluster~\cite{Snitch_2021}, a RISC-V-based \ac{pmca} optimized for energy-efficient \ac{fp} computation. It consisted of nine in-order RV32IMAF Snitch cores, each featuring a 32-bit SIMD-capable, mixed-precision \ac{fpu}. Two key ISA extensions, \ac{fp} Repetition~\cite{Snitch_2021} and Stream Semantic Registers~\cite{SSR_2021} enhanced execution efficiency by automating \ac{fp} loops and reducing memory overheads, enabling pseudo-dual-issue performance and achieving up to $\sim$90\% \ac{fpu} utilization on dense workloads. The cluster architecture included a shared L1 instruction cache and a 128KiB interleaved \ac{tcdm} connected via a single-cycle interconnect. One of the cores managed the \ac{dma} unit~\cite{DMA2023}, while the remaining eight executed parallel computations. We integrated the RedMulE~\cite{Redmule2022} accelerator to the cluster, an open-source, parametrizable accelerator optimized for reduced-precision \acp{MVM} using \ac{fp}16 and \ac{fp}8. We configured RedMulE to have 32 fused-multiply-accumulate blocks to minimize the final area of the cluster. We performed cycle-accurate \ac{rtl} simulations to obtain the performance metrics on the \ac{pmca}.

\section*{Results}\label{sec:results}
\subsection*{Accuracy Validation}\label{sec:accuracy_study}

We first validate the accuracy of the proposed method and compare it to conventional \ac{hwa} training~\cite{Rajendran75}. Table \ref{tab1} presents the performance validation of \emph{\ac{hwa}-\ac{lora} training} applied to MobileBERT on the SQuAD v1.1 dataset under different drift durations.
MobileBERT was chosen due to its relatively small size, making it practical for deployment on current \ac{aimc} chips~\cite{Le_Gallo_2023, Wen2024}. We will validate the effectiveness of our method on larger models in later sections.
Evaluation metrics used in the experiments include the F1 score and Exact Match (EM). The baseline results in the table represent the performance of the digital model without hardware constraints during training or inference.

The results demonstrate that \emph{\ac{hwa}-LoRA training} achieves performance comparable to conventional \ac{hwa} training, with F1 and EM scores within 1\% of those obtained through full \ac{hwa} training. Notably, at a drift time of 10 years, our approach outperforms the previous \ac{hwa} training method, achieving an F1 score of 85.36 versus 85.14, and an EM score of 76.92 versus 76.40. We attribute this improvement to the difference in update mechanisms: standard \ac{hwa} training updates all model parameters, which may cause the model to deviate significantly from the local minima reached during pretraining. In contrast, \emph{\ac{hwa}-LoRA training} updates only the low-rank components, helping the model stay closer to the flatter local minimum found during pretraining, which improves robustness under significant drift.

It is also worth noting that \emph{\ac{hwa}-LoRA training} updates only the LoRA weights, which make up approximately 6.6\% of the model parameters. Despite this small proportion of trainable parameters, the model demonstrates strong robustness to hardware constraints, as indicated by F1 and EM scores that are on par with conventional \ac{hwa} training. This suggests that adapting a model to hardware constraints may not require the adaption of all parameters. Instead, hardware adaptation could be inherently a low-rank problem, where LoRA-style updates are sufficient. Since hardware adaptation is a critical challenge in \ac{aimc}, these findings offer new insights and indicate that full model retraining may not always be necessary.

\begin{table}[h]
    \centering
    \caption{\textbf{Comparison of \ac{hwa} training and \ac{hwa}-\ac{lora} training}. The dataset used is SQuAD v1.1 and the model is MobileBERT.}

    \begin{tabular}{lccccccccc} 
        \toprule
        \multirow{2}{*}{\textbf{Training Method}} & \multirow{2}{*}{\textbf{Metric}} & \multirow{2}{*}{\textbf{Baseline}} & \multicolumn{7}{c}{\textbf{Score after Conductance Drift}} \\ 
        \cmidrule(l){4-10}
        & & & \textbf{0s} & \textbf{1h} & \textbf{1d} & \textbf{1w} & \textbf{1m} & \textbf{1y} & \textbf{10y} \\ 
        \midrule
        \multirow{2}{*}{\textbf{\ac{hwa} Training}} 
        & F1 & 90.01 & 89.47 & 89.07 & 88.66 & 88.19 & 87.73 & 86.59 & 85.14 \\
        & EM & 82.60 & 82.42 & 81.75 & 81.28 & 80.60 & 79.94 & 78.41 & 76.40 \\ 
        \midrule
        \multirow{2}{*}{\textbf{\ac{hwa}-\ac{lora} Training}} 
        & F1 & 89.17 & 89.06 & 88.71 & 88.36 & 87.97 & 87.49 & 86.51 & 85.36 \\
        & EM & 82.06 & 81.93 & 81.41 & 80.94 & 80.39 & 79.77 & 78.41 & 76.92 \\ 

        \bottomrule
    \end{tabular}
    \label{tab1}
\end{table}

\subsection*{Training Performance Evaluation}
Table \ref{tab:performance} benchmarks the number of trainable parameters and \ac{GPU} memory usage during training. The results show that \emph{\ac{hwa}-LoRA training} reduces the number of trainable parameters to approximately one million, which is more than a 15× reduction compared to conventional \ac{hwa} training. Additionally, \ac{GPU} memory usage is reduced by 13\%, saving over 4GB of VRAM, making training more accessible. Although MobileBERT only has 24.67M trainable parameters, the memory usage during \ac{hwa} training is significantly higher than during standard digital training due to the high overhead of modeling hardware constraints in both the forward and backward passes. This suggests that memory overhead is a key bottleneck in scaling \ac{aimc} to larger transformer models. \ac{lora} offers a promising solution to reduce both trainable parameters and memory requirements. Moreover, optimizing the training framework can further improve training efficiency. For instance, recent work using Triton to reduce memory usage and accelerate \ac{hwa} training presents another promising direction~\cite{buchel2024aihwkit}.

We further break down the effects of \ac{lora} placement and rank. Applying \ac{lora} only to the \ac{FFN} achieves lower parameter counts and memory usage than full \ac{hwa}-LoRA training, while restricting adaptation to the QKV projections reduces parameters even further (0.22M) with the lowest memory footprint (28.02 GB). For varying LoRA ranks ($r=1,2,4,8,16$), the number of parameters scales nearly linearly with $r$, whereas \ac{GPU} memory usage remains largely unchanged.

\begin{table}[H]
    \centering
    \caption{\textbf{Comparison on trainable parameters and \ac{GPU} memory usage across different training methods}. The model used is MobileBERT and the dataset is SQuAD v1.1. Experiments are conducted on an NVIDIA H100 80GB \ac{GPU} with a batch size of 32.}

    % \resizebox{\linewidth}{!}{%
    \begin{tabular}{lcc}
        \toprule
        Method & Trainable Parameters (M) & \ac{GPU} Memory Usage (GB)\\
        \midrule
        \ac{hwa} & 24.67 & 37.72\\ 
        \ac{hwa}-\ac{lora} & 1.63 & 32.92\\
        \ac{hwa}-\ac{lora} (FFN) & 1.40 & 31.88\\
        \ac{hwa}-\ac{lora} (QKV) & 0.22 & 28.02\\
        \midrule
        \ac{hwa}-\ac{lora} (r = 1) & 0.20 & 32.90\\
        \ac{hwa}-\ac{lora} (r = 2) & 0.41 & 32.90\\
        \ac{hwa}-\ac{lora} (r = 4) & 0.82 & 32.91\\
        \ac{hwa}-\ac{lora} (r = 8) & 1.63 & 32.92\\
        \ac{hwa}-\ac{lora} (r = 16) & 3.27 & 32.94\\
        
        \bottomrule
    \end{tabular}%
    % }
    \label{tab:performance}
\end{table}

\subsection*{Multi-Task Inference Evaluation}

\begin{table}[h]
    \centering
\caption{\textbf{\ac{hwa}-\ac{lora} training on the GLUE benchmark for  MobileBERT}. }
% \resizebox{\linewidth}{!}{%
\begin{tabular}{lcccccccc} 
\toprule
\multirow{2}{*}{\begin{tabular}[c]{@{}l@{}}\textbf{GLUE}\\\textbf{Task}\end{tabular}} & \multirow{2}{*}{\textbf{Score}} & \multicolumn{7}{c}{\textbf{Score after Conductance Drift}} \\ 
\cmidrule(l){3-9}
 &  & \textbf{0s} & \textbf{1h} & \textbf{1d} & \textbf{1w} & \textbf{1m} & \textbf{1y} & \textbf{10y} \\ 
\midrule
\textbf{SST-2} & 92.8 & 91.2 & 90.9 & 90.8 & 90.6 & 90.6 & 90.4 & 90.0 \\
\textbf{MNLI-m/mm} & 83.3/82.6 & 82.8/83.3 & 82.6/83.0 & 82.4/82.7 & 82.1/82.6 & 82.0/82.4 & 81.7/82.0 & 81.2/82.6  \\
\textbf{MRPC} & 88.8 & 87.6 & 87.5 & 87.5 & 87.2 & 87.3 & 87.1 & 86.4 \\
\textbf{QNLI} & 90.6 & 90.9 & 90.9 & 90.8 & 90.6 & 90.3 & 90.0 & 89.5 \\
\textbf{QQP} & 70.2 & 86.9 & 87.0 & 86.9 & 86.7 & 86.6 & 86.4 & 86.2 \\
\textbf{RTE} & 66.2 & 52.9 & 53.4 & 54.3 & 55.4 & 54.2 & 53.3 & 53.4 \\
\textbf{STS-B} & 84.4 & 87.5 & 87.3 & 87.0 & 86.9 & 86.7 & 86.2 & 85.6 \\
\textbf{CoLA} & 50.5 & 46.0 & 44.6 & 42.7 & 40.6 & 39.8 & 39.1 & 37.3 \\
\midrule
\textbf{GLUE} & 78.8 & 78.8 & 78.6 & 78.3 & 78.1 & 77.8 & 77.4 & 76.9 \\
\bottomrule
\end{tabular}
    \label{tab2}
\end{table}

We also evaluate our method on multi-task inference. Conventional \ac{hwa} methods require at least $N$ \ac{aimc} chips to handle $N$ tasks. In contrast, our proposed \emph{\ac{hwa}-LoRA training} method employs an elegant strategy wherein a single analog model is mapped to one \ac{aimc} chip with multiple sets of \ac{lora} weights, each adapted to a specific task. 

We report the results on the 8 tasks from the GLUE benchmark in Table~\ref{tab2}. These tasks were achieved using a single analog model mapped to \ac{aimc} tiles, with 8 sets of \ac{lora} parameters computed on \acp{DPU}, each containing 1.6M parameters, one for each task. To support all 8 tasks, a total of $8 \times 1.6$M \ac{lora} parameters are required, in addition to the 20.4M mappable parameters on \ac{aimc} tiles and 4.9M unmappable parameters on \acp{DPU}, resulting in a total of 38.1M parameters. In contrast, a conventional \ac{hwa} training approach~\cite{Rajendran75} would require developing and programming 8 separate models into the hardware, requiring at least $(8 \times 20.4 + 4.9)$M parameters. Our method, therefore, achieves a more than 4-fold reduction in the number of parameters.
Furthermore, our approach demonstrates robustness to the inherent constraints of \ac{aimc} hardware, with most tasks showing minimal performance degradation over time due to drift.

\emph{\ac{hwa}-LoRA training} also offers significant advantages, including on-chip task switching and on-chip adaptation to user data. For example, when \ac{aimc} hardware is initially configured for the SST-2 task, it can be easily switched to the MNLI task by simply updating the 1.6M \ac{lora} weights from SST-2 to MNLI, without needing to reload \ac{aimc} weights.
Additionally, the \ac{lora} weights allow for on-chip adaptation; when new user-generated data becomes available, the \ac{lora} weights can be updated accordingly, while the pre-trained model on the \ac{aimc} tiles remains unchanged.

\subsection*{Optimal Allocation of LORA Parameters}
Optimizing the allocation of \ac{lora} parameters is crucial for balancing model performance and resource efficiency. This challenge represents a typical resource allocation problem \cite{lammie2024lionheart}. While several methods exist for optimizing \ac{lora} allocation \cite{he2021towards, zhang2023adalora}, our study evaluates two predefined strategies for the sake of simplicity and consistency, avoiding the use of adaptive \ac{lora} allocation techniques.

As shown in Figure~\ref{fig3}a, increasing the LoRA rank generally improves the F1 score across all drift durations, though the gains eventually plateau. For instance, raising the \ac{lora} rank from 1 to 2 results in an F1 score increase of approximately 2.2 for a 10-year drift. However, increasing the rank from 8 to 16 yields only a marginal improvement of 1.0. Since a higher rank introduces greater computational overhead, this diminishing return suggests that selecting an appropriate rank can effectively balance accuracy gains with efficiency. Specifically, we find that a rank of 8 represents an optimal sweet spot for practical AIMC constraints: it provides sufficient capacity to compensate for analog noise and drift (accuracy constraint) while remaining lightweight enough to fit within the memory and latency budget of the digital accelerators (hardware constraint).

Figure \ref{fig3}b investigates the impact of different \ac{lora} allocation strategies on model accuracy over varying drift times. The results indicate that applying \ac{lora} to all linear layers reaches an initial accuracy of 89.06 (at 0 seconds), then shows a gradual performance decrease over time, reaching 85.36 after 10 years. This strategy has the highest \ac{lora} parameter count (1.6M). In contrast, restricting \ac{lora} to only the QKV linear layers or \ac{FFN} layers reduces the parameter count to 0.2M and 1.4M, respectively, but results in consistently lower F1 scores across all drift times. These findings suggest that applying \ac{lora} to all linear layers is essential for maintaining competitive F1 scores.

\begin{figure}[h]
\centering
\includegraphics[width=\columnwidth]{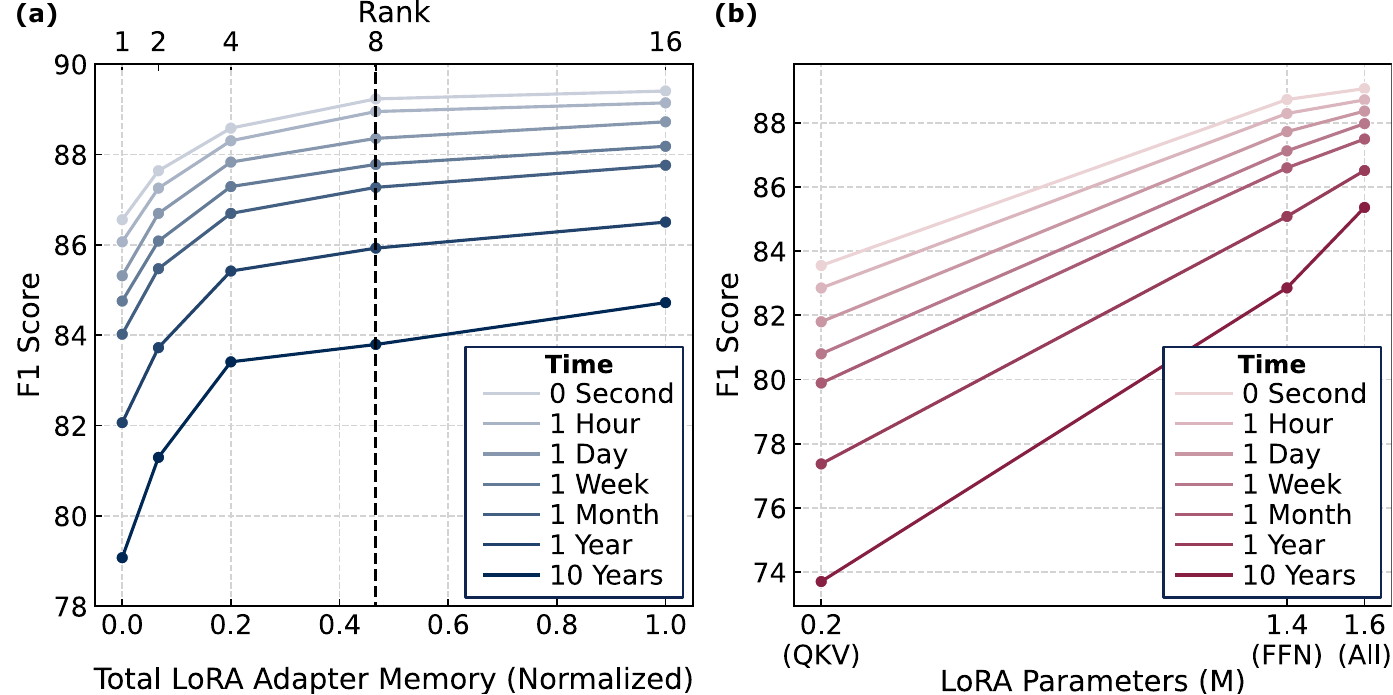} 
\caption{\textbf{Resource optimization study for MobileBERT and SQuADv1.1 at different drift times}. \textbf{a} Pareto front of the F1 score and total LoRA adapter memory for different rank values. A rank of 8 -- highlighted using a vertical dashed line -- provides a good tradeoff between these quantities.
\textbf{b} The total number of LoRA parameters when adapters are applied to different layers of the multi-head attention block.}
\label{fig3}
\end{figure}

\subsection*{Dynamic Adaptation}
In conventional approaches, once a model is deployed, its weights remain fixed, limiting adaptability to evolving hardware conditions. Our proposed \emph{\ac{hwa}-LoRA training} enables dynamic adaptation, which is crucial for real-world applications. For instance, as shown in Figure~\ref{fig4}a, reducing the \ac{ADC}/\ac{DAC} precision from 8-bit to 6-bit results in a significant 25\% F1 score drop after a 10-year drift period. Our approach addresses this issue by updating only the \ac{lora} weights.
This on-chip adaptation effectively mitigates performance degradation, improving the F1 score from 60.81 to 74.23 after ten years. Additionally, our method can adapt to other environmental variations, such as noise pattern shifts due to temperature fluctuations or changes in \ac{ADC} reference current, by leveraging \ac{lora} modules for targeted adjustments.

\begin{figure}[h]
\centering
\includegraphics[width=\columnwidth]{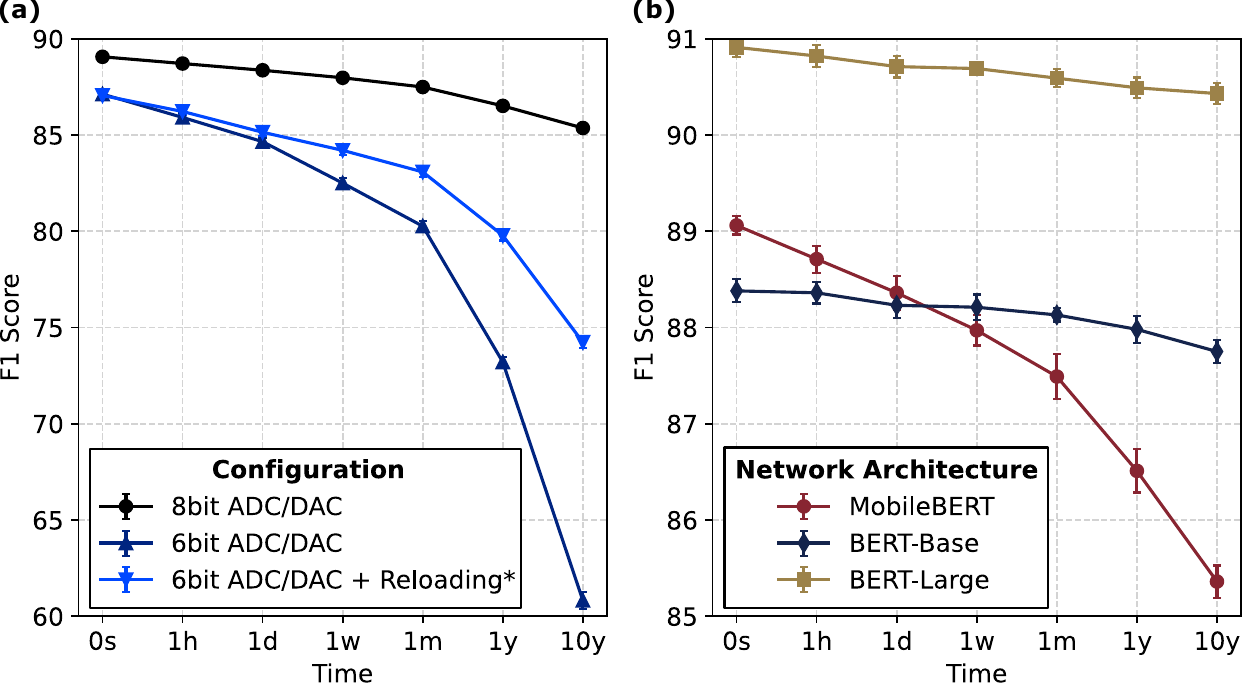} 
\caption{\textbf{
Dynamic adaptation and scalability studies at different drift times.}
\textbf{a} Different configurations for \ac{hwa}-\ac{lora} training evaluated on MobileBERT and SQuAD v1.1. *LoRA weight reloading. \textbf{b} Scalability analysis using BERT-Base and BERT-Large.}
\label{fig4}
\end{figure}

\subsection*{Scalability Studies}
As illustrated in Figure~\ref{fig4}b, our method scales effectively to both BERT-Base (108M parameters, with 1.3M \ac{lora} weights) and BERT-Large (334M parameters, with 3.5M \ac{lora} weights), optimizing these models for \ac{aimc} hardware by training only the \ac{lora} components. The results demonstrate that larger models not only achieve higher F1 scores but also exhibit increased robustness against hardware-induced degradation. For instance, a 10-year hardware drift leads to a performance drop of nearly 4 points in general; however, BERT-Base experiences only a 0.63-point reduction, while BERT-Large sees just a 0.48-point drop. This indicates that larger models are more resilient to \ac{aimc} hardware limitations, highlighting \ac{aimc}’s potential to support transformer-based models at scale.

Furthermore, BERT-Large is approximately  \unit[12]{$\times$} larger than MobileBERT, yet 
the \ac{lora} parameter count increases by only about \unit[2]{$\times$}, and a \ac{lora} rank of 8 remains sufficient. This highlights that our proposed method scales efficiently with minimal overhead. Encouraged by these results on encoder-based transformers, we next explore the application of our approach to decoder-only transformers, such as Large Language Models (LLMs).

\subsection*{Instruction Tuning and Reinforcement Learning on LLMs}
While previous sections have focused on encoder-only transformers, primarily due to their practical size for current \ac{lora} hardware, we additionally demonstrate the broader applicability of our \emph{\ac{hwa}-LoRA training} approach to decoder-only LLMs. Specifically, we investigate the LLaMA-3.1 model, containing 8 billion parameters, with a \ac{lora} rank of 16, comprising approximately 0.52\% of the model's total parameters. Detailed experimental configurations are provided in the Methods.

We first evaluate our approach in the context of instruction tuning using the Alpaca dataset, as shown in Table~\ref{tab4}. The baseline digital Llama 3.1 8B achieves strong zero-shot accuracy across diverse benchmarks, whereas the analog counterpart prior to \emph{\ac{hwa}-LoRA training} suffers severe degradation, with performance drops exceeding 40\% on several tasks. This substantial loss is primarily attributed to analog hardware noise, which affects reasoning and knowledge-retrieval capabilities in decoder-only LLMs. Notably, applying our \ac{hwa}-\ac{lora} training consistently restores a large portion of the lost performance, yielding absolute improvements of up to 38.23 percentage points (e.g., on HellaSwag) compared to the pre-tuning analog model. These results highlight that our method enables the model to recover general-purpose instruction-following abilities, demonstrating robust cross-task generalization without requiring full-precision retraining.

We extend our evaluation to reinforcement learning using the GSM8K dataset, aiming to enhance the model's reasoning capability through a long chain-of-thought (CoT), as shown in Table~\ref{tab5}. The digital baseline benefits from \ac{lora} fine-tuning, improving accuracy from 67.63\% to 85.06\%. In the analog setting, the unadapted model starts at a much lower accuracy of 37.98\%, reflecting its difficulty in conducting reasoning under hardware noise. After applying our\emph{\ac{hwa}-LoRA training}, the analog model reaches 70.74\% accuracy, representing a remarkable 32.76 percentage point gain and narrowing the gap to the digital post-\ac{lora} model by over half. These results show the versatility of \emph{\ac{hwa}-LoRA training}, extending its benefits beyond supervised instruction tuning to reinforcement learning, and demonstrating its ability to optimize models for complex, reasoning-intensive tasks.

\begin{table}[ht]
\centering
\begin{tabular}{lccccccccc}
\toprule
\textbf{Model Variant} & \textbf{Hella\-Swag} & \textbf{BoolQ} & \textbf{PIQA} & \textbf{Wino\-Grande} & \textbf{ARC-c} & \textbf{ARC-e} & \textbf{SciQ} & \textbf{COPA} & \textbf{OpenBookQA} \\
\midrule
Baseline: LlaMA 3.1 8B (Digital)  & 78.91 & 82.11 & 81.23 & 73.88 & 53.41 & 81.10 & 94.60 & 87.00 & 44.80 \\
Analog LlaMA 3.1 8B (Pre-\ac{hwa}-\ac{lora}) & 29.48 & 48.35 & 50.44 & 51.78 & 25.34 & 29.42 & 42.10 & 55.00 & 26.20 \\
Analog LlaMA 3.1 8B (Post-\ac{hwa}-\ac{lora}) & 67.71 & 69.36 & 69.64 & 55.80 & 35.67 & 54.71 & 80.10 & 71.00 & 33.60 \\
\bottomrule
\end{tabular}
\caption{\textbf{Zero-shot accuracy (\%) of LlaMA 3.1 8B across nine benchmarks under three settings:} 
(1) baseline digital model, (2) analog model before \emph{\ac{hwa}-\ac{lora} training}, and (3) analog model after \emph{\ac{hwa}-\ac{lora} training}. 
The analog model shows substantial accuracy degradation compared to the digital baseline. 
Applying \emph{\ac{hwa}-\ac{lora} training} substantially narrows this gap, yielding consistent improvements across all tasks. 
Additional results in the Supplementary Information show that evaluation at lower noise levels can further boost accuracy, bringing it closer to the digital baseline.}
\label{tab4}
\end{table}

\begin{table}[ht]
\centering
\begin{tabular}{lcccc}
\toprule
& \multicolumn{2}{c}{\textbf{Digital (LlaMA 3.1 8B)}} & \multicolumn{2}{c}{\textbf{Analog (LlaMA 3.1 8B)}} \\
\cmidrule(lr){2-3}\cmidrule(lr){4-5}
\textbf{Benchmark} & \textbf{Pre-\ac{lora}} & \textbf{Post-\ac{lora}} & \textbf{Pre-\ac{hwa}-\ac{lora}} & \textbf{Post-\ac{hwa}-\ac{lora}} \\
\midrule
GSM8K & 67.63 & 85.06 & 37.98 & 70.74 \\
\bottomrule
\end{tabular}
\caption{\textbf{GSM8K evaluation accuracy (\%) with chain-of-thought (CoT) reasoning.}  
The LLM is trained using reinforcement learning to produce outputs in the following format: \texttt{<start\_working\_out>} reasoning steps \texttt{<end\_working\_out>} \texttt{<SOLUTION>} final answer \texttt{</SOLUTION>}. We compare the digital baseline (\emph{LLaMA 3.1 8B}) and its analog counterpart, both before and after reinforcement learning via \ac{lora} training. In the digital case, \ac{lora} training improves performance from 67.63\% to 85.06\%. For the analog model, \emph{\ac{hwa}-\ac{lora} training} boosts accuracy from 37.98\% to 70.74\%, reducing the analog--digital performance gap from nearly 30\% to about 15\%. Additional results in the Supplementary Information show that, under reduced evaluation noise, this gap can shrink to 2.5\%.}

\label{tab5}
\end{table}

\subsection*{Latency Analysis of Hardware Components and Network Layers}
We analyze the performance optimization of the proposed hybrid implementation, which utilizes \ac{aimc} tiles and \acp{pmca}, with a focus on balancing their latencies. If the latency of the \acp{pmca} is well balanced with that of the \ac{aimc} tiles, an efficient \ac{aimc}-\ac{pmca} pipeline parallelism can be implemented, leading to minimal performance loss due to the additional computations introduced by the \ac{lora} modules. Then, we can enjoy the high accuracy on long drift time, efficient multi-task processing ability, and dynamic adaptation ability with minimal latency cost.
We examine this balance by analyzing our architecture across different layer sizes of MobileBERT, with the rank of \ac{lora} matrices fixed at 8.
We consider \ac{aimc} tile integration times of 128, 256, and \unit[512]{ns}, based on values reported in the literature for \ac{aimc}-based accelerators~\cite{Le_Gallo_2023}.
Since processing one token at a time using a \ac{pmca} does not fully exploit the parallelization capabilities of the cluster, we evaluate scenarios where multiple tokens are processed in parallel. Specifically, once the required static $XW$ (i.e., activation$\times$weight) \acp{MVM} on \ac{aimc} tiles are executed for $t$ tokens and transferred to the \ac{pmca}, the \ac{pmca} then performs the required $XAB$ \ac{MVM} \ac{lora} computations and element-wise addition operations. We consider $t$ values of 8, 16, 32, 64, and 128.

\begin{figure}[h]
\centering
\includegraphics[width=\columnwidth]{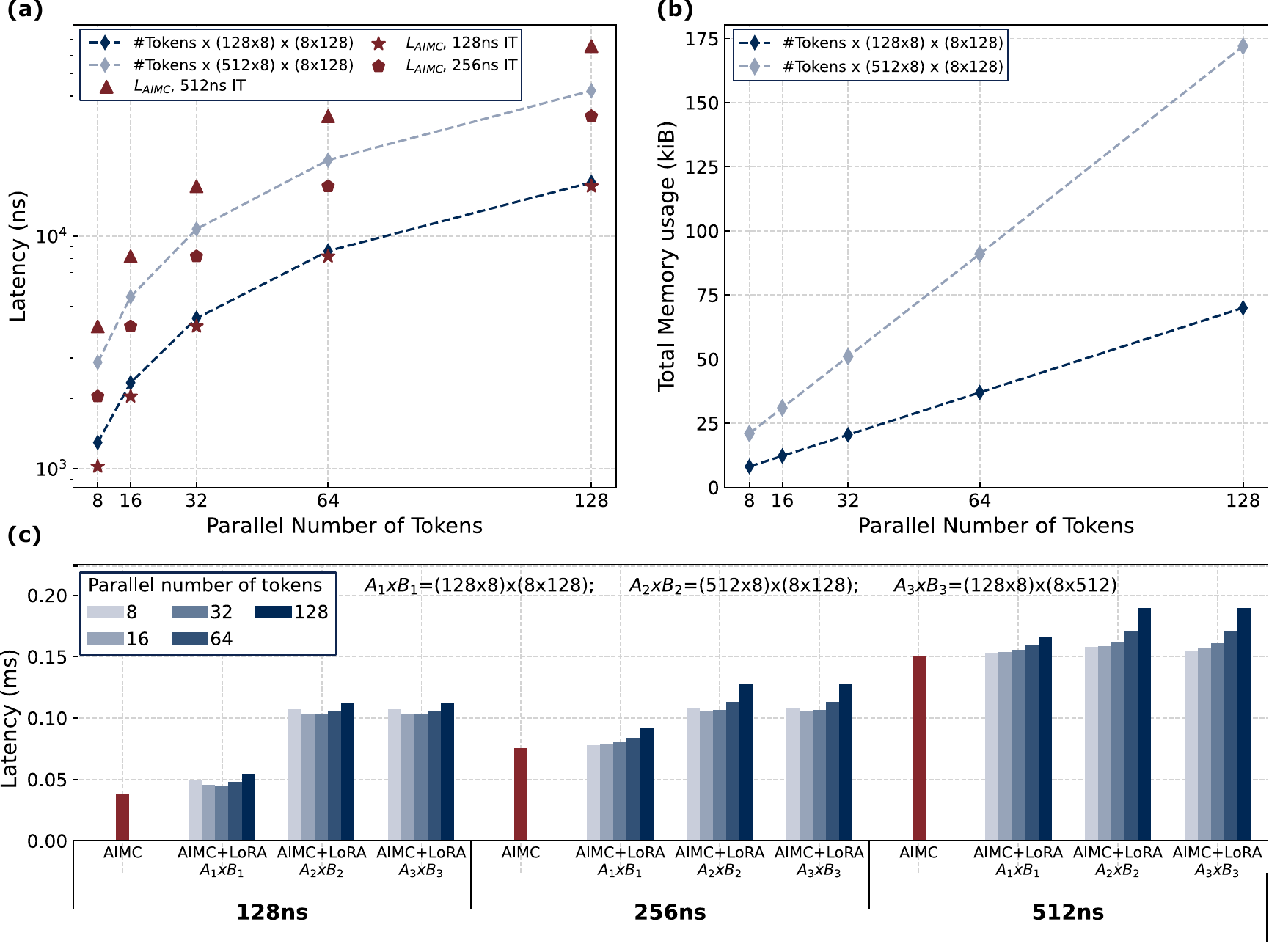} 
\caption{\textbf{Performance analysis and latency balancing of the proposed architecture with \ac{aimc} tiles and coupled \acp{pmca}}.
\textbf{a} The latency of two different MobileBERT layers with varying \ac{aimc} tile integration times and \textbf{b} \ac{pmca}’s TCDM requirement as a function of the number of parallel tokens. 
\textbf{c} The total latency for all the different layers of MobileBERT with an optimized \ac{aimc}-\ac{pmca} pipeline balancing the latency of the \ac{aimc} tiles and \acp{pmca}. 
The latency without any LoRA adapters (\ac{aimc}) is also reported.}
\label{fig:hw_study}
\end{figure}

The first step in our evaluation consists of analyzing and comparing the performance of \ac{aimc} tiles and \acp{pmca} when generating different numbers of parallel tokens, $t$.
Figure~\ref{fig:hw_study}a shows the latency of \ac{aimc} tiles and \acp{pmca} for different layer sizes, \ac{aimc} tile integration times, and $t$ values.
The results show that for a larger weight matrix size (512x128) and longer integration times (\unit[256]{ns} and \unit[512]{ns}), the latencies of \ac{pmca} and \ac{aimc} are well-balanced. For the same matrix size with a shorter integration time of \unit[128]{ns}, \ac{pmca} latency becomes the bottleneck.
In the case of smaller weight matrix size (128x128) and longer integration times (\unit[256]{ns} and \unit[512]{ns}), \ac{aimc} latency dominates, whereas for shorter integration time (\unit[128]{ns}) \ac{pmca} and \ac{aimc} latencies are balanced.
For a smaller weight matrix size (128x128), an optimal balance between \ac{pmca} and \ac{aimc} computation latencies is achieved when processing 128, 8, and 8 parallel numbers of tokens $t$ for integration times of 128, 256, and \unit[512]{ns}, respectively. This corresponds to \ac{pmca}-to-\ac{aimc} latency ratios of approximately 1.04, 0.63, and 0.32, respectively.
For the larger matrix size (512x128), the optimal balance is achieved for 128, 128, and 8 parallel tokens, corresponding to \ac{pmca}-to-\ac{aimc} ratios of approximately 2.57, 1.29, and 0.7 for the above respective integration times. 

Increasing the number of parallel tokens can enhance computational balance, but it also affects memory requirements. Figure~\ref{fig:hw_study}b shows the memory requirements to store the inputs and low-rank weight updates in the \acp{pmca} as a function of the number of parallel tokens $t$. We observed that the required \ac{pmca} memory varies significantly depending on the number of parallel tokens. For smaller weight matrices, the memory requirement ranges from \unit[8.2]{KiB} to \unit[21]{KiB}, while for larger matrices, it increases to between \unit[70]{KiB} to \unit[172]{KiB}. Considering the \acp{pmca} in our architecture have a \ac{tcdm} size of \unit[128]{KiB}, performing \acp{MVM} with large weight matrices and many parallel tokens requires either a larger \ac{tcdm} or additional \ac{tcdm}-SRAM communication.

To further assess the impact of the proposed methodology, we evaluate the total \ac{MVM} latency for processing a SL of 320 tokens. Figure ~\ref{fig:hw_study}c shows the total latency for different MobileBERT layers across different \ac{aimc} tile integration times. It compares the performance of the proposed \ac{hwa} \ac{lora} approach against a baseline \ac{aimc} system without \ac{lora} integration. We implement the \ac{aimc}-\ac{pmca} pipeline while accounting for data transfer latency between \ac{aimc} and \ac{pmca}.
As shown in Figure~\ref{fig:hw_study}c, the layer size, \ac{aimc} integration time, and the parallel number of tokens, all impact the performance overhead introduced by the additional \ac{lora} matrices. 
However, regardless of the integration time, when \ac{aimc} and \ac{pmca} latencies are well balanced, the additional latency from the \ac{lora} matrices remains minimal.
For large weight matrices (512$\times$128), the latency overhead from adopting \ac{hwa} \ac{lora} remains limited to \unit[2.72]{$\times$}, \unit[1.39]{$\times$}, and \unit[1.05]{$\times$} for integration times of 128, 256, and \unit[512]{ns}, respectively. Notably, in the best scenario, our method incurs only a \unit[4]{\%} latency overhead compared to the \ac{aimc} baseline. This evaluation highlights that balancing \ac{aimc} and \ac{pmca} latencies minimizes the overall layer latency and keeps \ac{hwa} \ac{lora}'s performance cost low.

This evaluation highlights the importance of balancing the \ac{aimc} and \ac{pmca} latencies. When the \ac{aimc} latency to perform the static \acp{MVM} aligns closely with the \ac{pmca} latency, the overall latency for each layer is minimized and the performance cost associated with the additional \ac{lora} matrices is kept minimal.

\section*{Discussion}
When deploying \acp{NN} on \ac{aimc} hardware, accuracy is often degraded due to hardware constraints. To mitigate this, researchers have introduced \ac{hwa} training, which adapts models to specific hardware characteristics. However, the trained models become tightly bound to the specific hardware settings and the dataset used during training, leading to over-specification. This limits both their flexibility and generalization ability.

One of the core motivations for using transformer models is their strong generalization capacity -- an advantage that is in direct conflict with the rigid nature of conventional \ac{aimc} deployment. Our proposed \emph{\ac{hwa}-LoRA training} method addresses this by retaining the meta-weights and introducing minimal overhead through low-rank adaptation. This design significantly improves generalization performance while maintaining accuracy comparable to conventional \ac{hwa} training methods. Furthermore, it enables additional capabilities for \ac{aimc}, such as multi-task inference and continual adaptability using a shared set of analog weights.

Importantly, our results provide concrete evidence that \ac{hwa}-LoRA training scales efficiently across model sizes and architectures. For encoder-only transformers, we evaluate MobileBERT (25M), BERT-Base (108M), and BERT-Large (334M), and observe that larger models exhibit smaller performance degradation under long-horizon drift (e.g., under 10-year simulated drift, MobileBERT drops by $\sim$4 points, while BERT-Base and BERT-Large degrade substantially less; see Table~\ref{tab1}). For decoder-only LLMs, we apply \ac{hwa}-LoRA training to LLaMA~3.1~8B for instruction tuning and reinforcement learning, where LoRA accounts for only 0.52\% of parameters, demonstrating that the adaptation overhead remains low even at billion-parameter scale. A plausible intuitive explanation is that larger models may possess stronger and more redundant representations, making them inherently more robust to perturbations and reducing the burden on the LoRA adapters to compensate hardware noise. Together, these experiments highlight that the same \ac{hwa}-LoRA training mechanism transfers cleanly across architectures (encoder/decoder), diverse task families (SQuAD/GLUE), model sizes (million-scale model/billion-scale model), and training paradigms (finetuning/RL).

The primary cost of our method lies in serving the LoRA components, whose overhead is proportional to their rank. Interestingly, we find that a rank of 8 is sufficient for most scenarios. This rank supports both learning a specific task and adapting to hardware variations -- indicating that the noise inherent in analog hardware with limited dynamic range can be effectively and efficiently compensated for. These positive results are expected to encourage broader adoption of analog hardware, allowing more users to benefit from its high speed and ultra-low power consumption.

Additionally, we found that the proposed architecture is well-suited for hardware implementations that parallelize \ac{aimc} tiles and \ac{pmca} for efficient execution. Our evaluation demonstrates that when the \ac{aimc} latency is well balanced with the \ac{pmca} latency, the latency overhead introduced by LoRA is minimal. In summary, our approach offers substantial improvements in the versatility and usability of \ac{aimc}, with only a modest latency cost.

For future improvements, insights from \ac{lora} and \ac{NN} training can be adapted to the \ac{aimc} context with suitable modifications~\cite{liu2024dora,jordan2024muon}. While this study builds on existing \ac{lora} methods, it also contributes back to that domain by demonstrating that \ac{lora} can be used to compensate for noisy weights—a scenario that was not considered in early \ac{lora} studies, which focused on full-precision~\cite{hu2022lora} or low-precision weights~\cite{dettmers2023qlora}, but not statistical (noisy) weights.

\ac{ADC} plays a crucial role in bridging the analog and digital domains, but its limited precision can impact overall system performance. Enhancing \ac{ADC} optimization to develop models resilient to lower bit-widths can accelerate inference or improve accuracy at fixed bit settings. Advances in transformer activation quantization are particularly relevant here~\cite{xiao2023smoothquant}, as they help mitigate the adverse effects of quantization noise. Notably, recent work by Büchel et al.~\cite{buchel2025analog} demonstrates that knowledge distillation can be used to overcome the precision limitations of \ac{ADC}, highlighting a promising direction for future research. Our proposed method is inherently compatible with such quantization and distillation techniques, making it well-suited for integration with these advances. Future work will incorporate these methods to enhance both performance and robustness in our \ac{aimc} system design.

Our initial results on instruction tuning and reinforcement learning provide promising insights into the broader applicability of the proposed methods and their scalability to \acp{llm}. We emphasize that the low-memory footprint of the proposed \ac{hwa}-\ac{lora} training is even more critical for \acp{llm} than for MobileBERT, as it enables training \acp{llm} with \ac{aimc} constraints on a single \ac{GPU}, making the process much more accessible to a wider range of researchers. Other benefits of \ac{hwa}-\ac{lora} training also extend naturally to \acp{llm}, including multi-task support through LoRA reloading or multi-LoRA serving, as well as dynamic adaptation to user-specific data.  A key challenge in this direction is to further improve the robustness of \ac{llm} training under hardware constraints. Our qualitative results in the Supplementary Information show that models can still generate coherent and meaningful outputs despite these hardware constraints. On the other hand, our quantitative results indicate that evaluation accuracy begins to degrade at a noise level of $4.0\%$ for instruction tuning and $2.0\%$ for reinforcement learning. A similar inflection point of $2$--$4\%$ was also reported by Buchel \textit{et al.}~\cite{buchel2025analog}. This indicates that there remains significant room for optimization in model training. Advancements in \ac{aimc} hardware would also contribute to this goal by enabling operation at lower noise levels.

Our results on GSM8K are closely related to the idea of noisy networks in reinforcement learning, which demonstrates that introducing controlled noise into neural networks can significantly enhance exploration capabilities and improve performance across both on-policy and off-policy methods. Similarly, the injection of noise in our method may encourage exploration. From this perspective, the inherent noise within \ac{aimc} is not a limitation to be overcome, but rather a beneficial feature that drives more effective learning. How to maximise this potential benefit can be an interesting future work.

The proposed method in this paper may also be connected to the idea of mortal computations \cite{ororbia2023mortal, hinton2022forward}, which involves using analog hardware to perform neural network operations.
In such systems, the algorithm is inherently tied to the analog substrate, meaning the computation is mortal -- it dies with the analog hardware.
While this tight integration enables high-speed and energy-efficient processing, it also presents challenges in training effectiveness, knowledge sharing, and scalability, as identified in a previous study \cite{hinton2022forward}.
Our method introduces minimal digital parameters (i.e., LoRA weights) to augment the analog weights components. This approach addresses learning limitations without requiring precise knowledge of the analog weights’ exact values; instead, it only relies on their statistical noise profiles, making the mortal system learnable. The LoRA modules effectively compensate for the noise in the mortal system, enabling robust functionality. As a result, the architecture achieves an elegant balance: analog weights remain fast and large in number, while the digital weights, though less-efficient and slower, are kept minimal. When adaptation is needed, only the minimal digital weights are updated. This update process is simpler and more efficient than modifying all analog weights.

\newpage

\section*{Acknowledgments}
This work is supported in part by the NeuroSoC project funded under Horizon Europe Grant Agreement 101070634. The work of Bipin Rajendran was supported in part by
EPSRC Open Fellowship under Grant EP/X011356/1 and in part by EPSRC under Grant EP/X011852/1. We also gratefully acknowledge valuable discussions with Dr. Abu Sebastian and help from Julian Büchel regarding the usage of AIHWKIT.  

\section*{Author contributions}
C.~L. proposed the initial concept of \ac{hwa} \ac{lora} training and performed the accuracy validation experiments. E.~F. conducted the latency analysis. C.~L. and E.~F. co-authored the initial draft of the manuscript, with all authors contributing substantial feedback and revisions. B.~R., M.~G., and I.~B. provided overall project supervision and guidance.

\newpage
\section*{References}
\bibliography{main}

% \clearpage
% \begin{figure*}
% \centering
% \includegraphics[width=\columnwidth]{Figures_last/fig4.jpg} 
% \caption{\textbf{
% Instruction tuninig and reinforcement learning on LLMs.}
% \textbf{a} Different configurations for \ac{hwa}-\ac{lora} evaluation on MobileBERT and SQuAD v1.1. *LoRA weight reloading. \textbf{b} Scalability analysis using BERT-Base and BERT-Large.}
% \label{fig4}
% \end{figure*}

% \clearpage
% \begin{figure*}
% \centering
% \includegraphics[width=\columnwidth]{Figures_last/fig5.jpg} 
% \caption{\textbf{
% Instruction tuninig and reinforcement learning on LLMs.}
% \textbf{a} Different configurations for \ac{hwa}-\ac{lora} evaluation on MobileBERT and SQuAD v1.1. *LoRA weight reloading. \textbf{b} Scalability analysis using BERT-Base and BERT-Large.}
% \label{fig5}
% \end{figure*}

% =====================================================
% SUPPLEMENTARY MATERIAL - TO BE INJECTED BEFORE \end{document}
% =====================================================

% Custom environments and formatting (if not already defined in main document)

% Example box environment
\newenvironment{examplebox}[2]{%
  \begin{tcolorbox}[title=Example #1 (#2)]
  \begin{description}
}{%
  \end{description}
  \end{tcolorbox}
}

% Fix table spacing
\setlength{\intextsep}{10pt plus 2pt minus 2pt}  % space around floats
\setlength{\textfloatsep}{10pt plus 2pt minus 2pt}  % space between floats and text

% =====================================================
% SUPPLEMENTARY CONTENT BEGINS
% =====================================================

\newpage
\section*{SUPPLEMENTARY NOTE 1: Ablation study on AHWA-LoRA training}
\label{sec:s1}

We conduct a comprehensive ablation study to analyze the impact of key hyperparameters in \emph{AHWA-LoRA training}: learning rate, weight noise, and clipping method. All experiments use MobileBERT on the SQuAD v1.1 dataset, with performance evaluated using F1 scores after simulating conductance drift over different time periods.

\subsection*{Learning Rate}

As shown in Table \ref{tab:lr_ablation}, a learning rate of $2 \times 10^{-4}$ provides the best overall results. A very low learning rate, such as $5 \times 10^{-6}$, results in a higher training loss and significantly lower F1 scores, indicating that the model fails to converge effectively. While increasing the learning rate to $5 \times 10^{-5}$ improves performance, it remains suboptimal compared to $2 \times 10^{-4}$. Conversely, a high rate of $8 \times 10^{-4}$ makes the training process unstable and causes the model to collapse.

\begin{table}[ht]
\centering
\caption{\textbf{Ablation study on learning rate.} Results are obtained on SQuAD v1.1 using MobileBERT with \emph{AHWA-LoRA training}.}
\label{tab:lr_ablation}
\begin{tabular}{lcccc}
\toprule
\multirow{2}{*}{\textbf{Metric}} & \multicolumn{4}{c}{\textbf{Learning Rate}} \\
\cmidrule(l){2-5}
& \textbf{5e-6} & \textbf{5e-5} & \textbf{2e-4} & \textbf{8e-4} \\
\midrule
\textbf{Train Loss} & 1.9505 & 1.2004 & 0.9306 & Collapse \\
\midrule
\multicolumn{5}{l}{\textbf{F1 Score after Conductance Drift}} \\
\midrule
\textbf{0s} & 73.49 & 86.64 & 89.06 & - \\
\textbf{1h} & 72.13 & 86.30 & 88.71 & - \\
\textbf{1d} & 71.07 & 85.74 & 88.36 & - \\
\textbf{1w} & 69.64 & 85.08 & 87.97 & - \\
\textbf{1m} & 68.08 & 84.48 & 87.49 & - \\
\textbf{1y} & 66.76 & 83.26 & 86.51 & - \\
\textbf{10y} & 63.27 & 81.41 & 85.36 & - \\
\bottomrule
\end{tabular}
\end{table}

\footnotetext{*20 epochs training, which is 5 epoches longer than others}

\subsection*{Weight Noise}

We investigate how varying the standard deviation of Gaussian weight noise during training affects model performance. As shown in Table \ref{tab:noise_ablation}, moderate noise levels achieve optimal results. The F1 score following conductance drift initially improves with increasing noise levels, reaching peak performance between 0.067-0.075 before declining. A noise level of 0.067 delivers the highest overall performance, while 0.075 with extended training (20 epochs) provides marginally better long-term stability, sustaining an F1 score of 86.49 after 10 years, though this requires additional training epochs. Lower noise levels result in more severe performance degradation over time and reduced robustness to drift. Conversely, excessive noise (0.12) completely destabilizes the training process, causing training failure.

\begin{table}[ht]
\centering
\caption{\textbf{Ablation study on weight noise.} Results are obtained on SQuAD v1.1 using MobileBERT with\emph{AHWA-LoRA training}.}
\label{tab:noise_ablation}
\begin{tabular}{lccccccc}
\toprule
\multirow{2}{*}{\textbf{Metric}} & \multicolumn{7}{c}{\textbf{Weight Noise Level}} \\
\cmidrule(l){2-8}
& \textbf{0.02} & \textbf{0.0377} & \textbf{0.067} & \textbf{0.075} & \textbf{0.075*} & \textbf{0.09} & \textbf{0.12} \\
\midrule
\textbf{Train Loss} & 0.5392 & 0.6876 & 0.9306 & 1.0223 & 0.9373 & 1.1836 & Collapse \\
\midrule
\multicolumn{8}{l}{\textbf{F1 Score after Conductance Drift}} \\
\midrule
\textbf{0s} & 87.46 & 88.74 & 89.33  & 88.74 & 89.18 & 87.39 & - \\
\textbf{1h} & 86.87 & 88.28 & 89.23  & 88.65 & 88.74 & 87.36 & - \\
\textbf{1d} & 85.50 & 87.59 & 88.28 & 88.29 &  88.37 & 86.98 & - \\
\textbf{1w} & 84.74 & 87.08 & 88.23 & 88.08 & 88.51 & 86.62 & - \\
\textbf{1m} & 83.26 & 85.56 & 87.86 & 87.71 & 87.98 & 86.37 & - \\
\textbf{1y} & 80.29 & 84.00 & 86.97 & 86.89 & 87.17 & 85.83 & - \\
\textbf{10y} & 74.97 & 80.63 & 85.76 & 85.68 & 86.49 & 85.17 & - \\
\bottomrule
\end{tabular}
\end{table}

\subsection*{Clipping Method}

Table \ref{tab:clipping_ablation} demonstrates that adaptive clipping at 2.5$\sigma$ of the weight distribution provides optimal results, achieving the highest F1 scores both initially (89.51) and after long-term drift (86.01 at 10 years). The 3$\sigma$ threshold used in our main experiments also performs well. However, overly restrictive clipping at 2$\sigma$ causes training collapse, while non-adaptive fixed clipping (Fixed 1) yields suboptimal performance, highlighting the importance of distribution-aware adaptive clipping strategies.

\begin{table}[ht]
\centering
\caption{\textbf{Ablation study on weight clipping method.}  Results are obtained on SQuAD v1.1 using MobileBERT with \emph{AHWA-LoRA training}.}
\label{tab:clipping_ablation}
\begin{tabular}{lcccc}
\toprule
\multirow{2}{*}{\textbf{Metric}} & \multicolumn{4}{c}{\textbf{Clipping Value}} \\
\cmidrule(l){2-5}
& \textbf{3$\sigma$} & \textbf{2.5$\sigma$} & \textbf{2$\sigma$} & \textbf{Fixed 1} \\
\midrule
\textbf{Train Loss} & 0.9306 & 0.8345 & Collapse & 1.0299 \\
\midrule
\multicolumn{5}{l}{\textbf{F1 Score after Conductance Drift}} \\
\midrule
\textbf{0s} & 89.06 & 89.51 & - & 88.57 \\
\textbf{1h} & 88.71 & 89.13 & - & 88.11 \\
\textbf{1d} & 88.36 & 88.80 & - & 87.93 \\
\textbf{1w} & 87.97 & 88.42 & - & 87.56 \\
\textbf{1m} & 87.49 & 88.29 & - & 87.28 \\
\textbf{1y} & 86.51 & 86.95 & - & 86.11 \\
\textbf{10y} & 85.36 & 86.01 & - & 84.32 \\
\bottomrule
\end{tabular}
\end{table}

\section*{SUPPLEMENTARY NOTE 2: Additional quantitative and qualitative  results for AHWA-LoRA training on instruction tuning}
\label{sec:supp_inst}

Table~\ref{tab4} reports the robustness of the analog LLaMA~3.1~8B model trained with \emph{AHWA-LoRA training} when evaluated under varying inference noise levels. The model was trained with a noise level of $6.7\%$. On the HellaSwag dataset, the analog model maintains accuracy above $75\%$ for noise levels up to $4.0\%$, with only a gradual degradation from $78.91\%$ at $0.0\%$ noise (digital baseline) to $75.34\%$ at $4.0\%$ noise. Performance declines more rapidly beyond $4.0\%$ noise, dropping to $70.76\%$ at $6.0\%$ noise and $67.71\%$ at the training noise level of $6.7\%$. When evaluated using the PCM model (no drift), it achieves $76.21\%$ accuracy.

\begin{table}[ht]
\centering
\caption{\textbf{LLaMA~3.1~8B evaluation on HellaSwag accuracy (\%) across different inference noise levels.} The digital (noiseless) baseline is included at $0.0\%$ noise for comparison. The analog model was trained with \emph{AHWA-LoRA training} (training noise level $6.7\%$).}
\label{tab4}
\renewcommand{\arraystretch}{1.2}
\setlength{\tabcolsep}{6pt}
\begin{tabular}{lccccccccc}
\toprule
\textbf{Noise level (\%)} 
& \textbf{Digital (0.0)} & \textbf{1.0} & \textbf{2.0} & \textbf{3.0} 
& \textbf{4.0} & \textbf{5.0} & \textbf{6.0} & \textbf{6.7} 
& \textbf{PCM (0s drift)} \\ 
\midrule
\textbf{HellaSwag Accuracy (\%)} 
& 78.91 & 77.65 & 77.18 & 76.54
& 75.34 & 73.36 & 70.76 & 67.71 & 76.21 \\
\bottomrule
\end{tabular}
\end{table}

\noindent\textbf{Qualitative Example (Instruction-Following).}  
Below is an example of the model's output for a simple instruction-only prompt after \emph{AHWA-LoRA training}. The response is concise, factual, and follows the instruction accurately.

\begin{quote}
\textbf{Instruction:} What is a famous tall tower in Paris?  

\textbf{Response:} One famous tall tower in Paris is the Eiffel Tower. The Eiffel Tower is an iron lattice tower constructed in 1889 and was the tallest building in the world at the time of its construction. It stands at a height of 1,063 feet (324 meters) and is one of the most recognizable landmarks in the world.
\end{quote}

\section*{SUPPLEMENTARY NOTE 3: Additional quantitative and qualitative results for AHWA-LoRA training on reasoning (GSM8K)}
\label{sec:supp_gsm8k}

Table~\ref{tab:gsm8k_noise_accuracy} presents GSM8K evaluation results for the analog LLaMA~3.1~8B model trained with \emph{AHWA-LoRA training} at a training noise level of $3.0\%$. The model exhibits strong robustness up to $2.0\%$ noise, with only a modest drop from $85.06\%$ (digital baseline) to $83.55\%$. At the training noise level ($3.0\%$), accuracy drops sharply to $70.74\%$, indicating reduced reliability under higher noise. Evaluation with the PCM model (no drift) yields an accuracy of $77.71\%$.

\begin{table}[ht]
\centering
\caption{\textbf{LLaMA~3.1~8B evaluation on GSM8K accuracy (\%) across selected noise levels.} The digital (noiseless) baseline is included at $0.0\%$ noise for comparison. The analog model was trained with \emph{AHWA-LoRA training} (training noise level $3.0\%$).}
\label{tab:gsm8k_noise_accuracy}
\renewcommand{\arraystretch}{1.2}
\setlength{\tabcolsep}{6pt}
\begin{tabular}{lccccc}
\toprule
\textbf{Noise level (\%)} 
& \textbf{Digital (0.0)} & \textbf{1.0} & \textbf{2.0} & \textbf{3.0} & \textbf{PCM (0s drift)} \\ 
\midrule
\textbf{GSM8K Accuracy (\%)} 
& 85.06 & 84.53 & 83.55 & 70.74 & 77.71 \\
\bottomrule
\end{tabular}
\end{table}

\noindent\textbf{Evaluation Samples (Chain-of-Thought Reasoning).}  
We present two examples from GSM8K (evaluation noise level $3.0\%$) showing the model's reasoning traces and final machine-readable answers.  
Each answer follows the format:
\begin{itemize}
    \item A free-form reasoning trace between \verb|<start_working_out>| and \verb|<end_working_out>|.
    \item A final answer enclosed in \verb|<SOLUTION>| and \verb|</SOLUTION>| tags.
\end{itemize}
This structure allows for automated correctness checks and could enable reinforcement learning with verifiable outputs, encouraging the model to think before answering.

\begin{examplebox}{1}{Incorrect}
  \item[Question] Josh decides to try flipping a house. He buys a house for \$80{,}000 and then spends \$50{,}000 on repairs. This increases the value of the house by \(150\%\). How much profit did he make?
  \item[Expected Answer] \(\mathbf{70{,}000}\)
  \item[Model Output]
\begin{verbatim}
<start_working_out>
Josh bought the house for $80,000 and added $50,000 in repairs.
The total spent is $80,000 + $50,000 = $130,000.
The value of the house increased by 150% of the original price:
1.5 * $80,000 = $120,000.
The profit is the new value minus the original price:
$130,000 - $80,000 = $50,000.
<end_working_out>
<SOLUTION>50000</SOLUTION>
\end{verbatim}
  \item[Extracted] \(50{,}000\)
  \item[Correct] \textcolor{red}{False}
\end{examplebox}

\begin{examplebox}{2}{Correct}
  \item[Question] A robe takes 2 bolts of blue fiber and half that much white fiber. How many bolts in total does it take?
  \item[Expected Answer] \(\mathbf{3}\)
  \item[Model Output] 
\begin{verbatim} 
<start_working_out>
The robe needs 2 bolts of blue fiber.
White fiber is half as much: 2 / 2 = 1 bolt.
Total = 2 + 1 = 3.
<end_working_out>
<SOLUTION>3</SOLUTION>
\end{verbatim}
  \item[Extracted] \(3\)
  \item[Correct] \textcolor{green}{True}
\end{examplebox}

\end{document}